\documentstyle{article}

\begin{document}

\begin{center}
{\huge\bf On Experimental Evidence of Uncertainty of Quantized  
Electromagnetic Potential from QHE}
\end{center}

\vspace{1cm}
\begin{center}
{\large\bf
F.GHABOUSSI}\\
\end{center}

\begin{center}
\begin{minipage}{8cm}
Department of Physics, University of Konstanz\\
P.O. Box 5560, D 78434 Konstanz, Germany\\
E-mail: ghabousi@kaluza.physik.uni-konstanz.de
\end{minipage}
\end{center}

\vspace{1cm}

\begin{center}
{\large{\bf Abstract}}
\end{center}
It is shown that the observed potential drops on QHE samples can be  
considered as a realization of uncertainty relation for the  
quantized two dimensional electromagnetic potential.
\begin{center}
\begin{minipage}{12cm}

\end{minipage}
\end{center}

\newpage
Recently it was shown that in accord with the {\it canonical} flux  
quantization, there should be an uncertainty relation for quantized  
two dimensional electromagnetic potential \cite{min}. This  
uncertainty relation which is comparable with the usual quantum  
mechanical uncertainty relations is of the form $e \Delta A_x \cdot  
\Delta x \geq \hbar$, where $e$, $\Delta A_x$ and $\Delta x$ are,  
respectively, the electron charge, the uncertainty of  
electromagnetic potential and the uncertainty of position of  
electrons. In the context of QHE the position uncertainty of  
electrons is the width wherein the edge current flows on the {\it  
edge} of sample. This width, which is in general larger than the  
length scale of magnetic length, is given for any QHE sample in  
accord with the experimental preparation of sample under quantum  
Hall conditions. In the "ideal" case the edge current flows close to  
the edge within the length scale of the respective magnetic length  
$l_B$ \cite{K}, which is determind by QHE data of the sample. Its  
value is defined by: $l_B ^2 := \displaystyle{\frac{\hbar}{e B}}$ or  
in the single electron picture of IQHE, by $\nu := 2 \pi n l_B ^2$,  
where $B$, $\nu$ and $n$ are, respectively, the applied magnetic  
field, the filling factor and the global density of electrons on the  
sample.

\bigskip
Here we report on the possibility that the so called potential  
drops on QHE samples \cite{all} can be considered as realizations of  
the mentioned uncertainty relation. Recall also that, although  
electromagnetic potential is not measurable in view of its gauge  
dependence, nevertheless poterntial differences are gauge invariant  
and measurable.

Generally, these experiments can be classified in two groups: One  
group reports on potential drops which appear on the edge of QHE  
samples on a width which is equal to the respective $(l_B)^{-1}$  
values [3a], [3b].

The other group reports on similar potential drops but on a width  
which is smaller than the respective $(l_B)^{-1}$ values [3c].

We consider the first type potential drops as the {\it maximal} and  
the second type as the {\it general} potential drops. Thus, we will  
show that potential drops are (quantum) uncertainty of  
electromagnetic potential in accord with the mentioned uncertainty  
relation: Whereby the maximal potential drops appear in cases, where  
in view of QHE preparation, the position uncertainty of electrons  
on the sample is the most minimal one which is equal to the magnetic  
length, i. e. $(\Delta x)_{(minimum)} = l_B$.
Whereas the general potential drops are electromagnetic potential  
uncertainties which appear in cases with a position uncertainty  
$\Delta x > l_B$.

\bigskip
Therefore, let us briefly sketch the quatum theoretical basis of  
this uncertainty relation.

The point of departure is the {\it canonical} conception of flux  
quantization $\Phi = \oint e A_m dx^m = \int \int e F_{mn} dx^m  
\wedge dx^n = {\mathbf Z} h$, where $A_m$ and $F_{mn}$ are the  
electromagnetic potential and the magnetic field strength and $m, n  
= 1, 2$. We proved that in this case there must be a new uncertainty  
relation $e \Delta A_m . \Delta x_m \geq \hbar$ which is related  
with the flux quantization $\Phi = \oint e A_m dx^m = {\mathbf Z} h$  
and with the quantum commutator postulate $e [ \hat{A}_m  ,   
\hat{x}_n ] = -i \hbar \delta_{mn}$ \cite{min}.

To see the consistency of this postulate let us remark that such a  
commutator can be considered as a result of electronic behaviour in  
magnetic fields in the following manner:

>From the usual requirement in flux quantization that the  
electronic current density $j_m = ne \hat{V}_m = \Psi^* (\hat{p}_m -  
e \hat{A}_m) \Psi$ must vanish in the region where the contour  
integral $\oint A_m dx^m$ takes place, one concludes that {\it in  
this region} $[ \hat{V}_m \ , \ \hat{x}_n ] = 0$. This implies that  
in this region $[\hat{p}_m \ , \ \hat{x}_n ] = e [ \hat{A}_m \ , \  
\hat{x}_n ]$ or that $ e [ \hat{A}_m s , \ \hat{x}_m ] = -i \hbar  
\delta_{mn}$.

\medskip
Moreover, it is also known from cyclotron motion of electrons, that  
the coordinate operators of relative coordinates are non-commuting  
[4a]. Thus, one has $[\hat{x}_m \ , \ \hat{x}_n] = -i l_B ^2  
\epsilon_{mn}$ for the relative cyclotron coordinates, where $l_B$  
is the magnetic length [4b]. Now this commutator is proportional to  
the mentioned commutator $[ \hat{A}_m \ , \ \hat{x}_n] = -i  
\delta_{mn} \hbar$ by the usual Landau gauge $A_m = B x^n  
\epsilon_{mn} \ , \epsilon_{mn} = - \epsilon_{nm} = 1$.

\bigskip
We showed also rigorously that the flux quantization can be  
understood as the canonical quantization of flux functional $\Phi =  
\oint e A_m dx^m = \int \int e F_{mn} dx^m \wedge dx^n$ on the phase  
space of flux system which contains the set of canonical conjugate  
variables ${\{ A_m \ , x^m }\}$. Thus, in accord with geometric  
quantization \cite{wood} the quantum differential operators on the  
quantized phase space of this system should be given by $\hat{A}_m =  
-i \hbar \displaystyle{\frac{\partial}{\partial x^m}}$ and   
$\hat{x}_m = i \hbar \displaystyle{\frac{\partial}{\partial A_m}}$  
\cite{min}. Recall however that the wave function of quantized  
${\{A_m , x^m}\}$ system should be considered either in the $\Psi (  
A_m , t)$- or in the $\Psi ( x^m , t)$ representation. Therefore,  
the quantum operators should be given, respectively, either by the  
set ${\{ \hat{A}_m = A_m \ , \ \hat{x}_m =
i \hbar \displaystyle{\frac{\partial}{\partial A_m}}}\}$ or by the  
set ${\{ \hat{A}_m = -i \hbar\displaystyle{\frac{\partial}{\partial  
x^m}} \ , \ \hat{x}^m = x^m }\}$ \cite{min}.

In both representations the commutator between the quatum operators  
is given by $(-i \hbar)$:

\begin{equation}
e [ \hat{A}_m \ , \ \hat{x}_n ] \Psi = -i \hbar \delta_{mn} \Psi
\end{equation}
\label{six}

Equivalently, we have in accord with quantum mechanics a true  
uncertainty relation for $A_m$ and $x_m$, i. e.: $e \Delta A_m \cdot  
\Delta x_m \geq \hbar$.

\bigskip
Furthermore, the electromagnetic gauge potential have in accord  
with the uncertainty relations

$e \Delta A_m \cdot \Delta x_m \geq \hbar$ a maximal uncertainty of  
$(\Delta A_m)_{(maximum)} = \displaystyle{\frac{\hbar}{e l_B}}$ for  
the case where the position uncertainty aquires its most minimal  
value. This "ideal" case where $(\Delta x_m)_{(minimum)} = l_B$  
corresponds to the "uncertainty equantions" $e \Delta A_m \cdot  
\Delta x_m = e B \Delta x_m \cdot \Delta x_n |\epsilon_{mn}| =  
\hbar$, from which one can obtain the independent definition of  
magnetic length $ l^2 _B = \displaystyle{\frac{\hbar}{e B}}$  
\cite{min}. This procedure proves the consistency of the approach.

\bigskip
Hence we show that the observed potential drops in QHE experiments  
which is reported in \cite{all} can be considered as experimental  
evidences for the above uncertainty relation.

\bigskip
>From topological point of view, which is useful in a topological  
effect like QHE, all usual two ddimensional QHE samples are  
equivalent to a disc. Thus, one should consider QHE on such a sample  
with the radial $(r)$ and azimuthal $(\phi)$ degrees of freedom,  
where $ ( {\{ x_m}\} \sim {\{ r \, , \, \phi }\} )$ and $( {\{ A_m  
}\} \sim {\{ A_r \, , \, A_{\phi} }\} )$.

The uncertainty relation $e \Delta A_m \cdot \Delta x_m \geq \hbar$  
asserts that the general potential uncertainty is given by $\Delta  
A_m \geq \displaystyle{\frac{\hbar}{e \Delta x_m}}$. If we identify  
the azimuthal or the {\it edge} component $A_{\phi}$ with Hall  
potential $A_H$, then the uncertainty relation for the Hall  
potential is given by $\Delta A_H \cdot \Delta x_H \geq  
\displaystyle{\frac{\hbar}{e}}$ where $\Delta x_H$ is the position  
uncertainty for electrons which flow in Hall current on the edge of  
sample. Thus, the general potential uncertainty for Hall potential  
is given in accord with the uncertainty relation by $\Delta A_H \geq  
\displaystyle{\frac{\hbar}{e \Delta x_H}}$, whereas the maximal  
potential uncertainty for $A_H$ is given by $(\Delta  
A_H)_{(maximum)} = \displaystyle{\frac{\hbar}{e l_B}}$ for the case  
$(\Delta x_H)_{(minimum)} = l_B$.

Now, in the first group, Ref. [3b] reports on observation of  
potential drops across the QHE samples over a width of $100 \; \mu  
m$ from the edge of samples.
This experiment is performed on QHE samples with filling factor  
$\nu = 2$ and a magnetic length value of $l_B = 10^{-2} \; \mu m$  
[3b].

On the other hand, the theoretical value for the maximal potential  
uncertainty for Hall potential which is obtained from uncertainty  
relation is $(\Delta A_H)_{(maximum)} = \displaystyle{\frac{\hbar}{e  
l_B}}$. Therefore, on obtains with $l_B = 10^{-2} \; \mu m$ the  
value $(\Delta A_H)_{(maximum)} \approx 100 \; \mu m$ \cite{dim}.

It agrees with the observed width of potential drop in Ref. [3b].

The other experiment in the first group [3a] is performed under  
almost the same QHE conditions as in Ref. [3b], but with filling  
factor $\nu = 4$. The value of magnetic length for this sample is  
obtained to be $l_B \approx 1.4 \cdot 10^{-2} \ \mu m$ in accord  
with QHE data in Ref. [3a]. Therefore, we obtain for the theoretical  
value of $(\Delta A_H)_{(maximum)} = \displaystyle{\frac{\hbar}{e  
l_B}}$ in this case $(\Delta A_H)_{(maximum)} \approx 70 \ \mu m$  
\cite{dim}. It agrees also with the observed value in Ref. [3a].

\bigskip
In this sense, from theoretical point of view, our maximal Hall  
potential uncertainty results which correspond to the respective  
most minimal position uncertainties or to the respective magnetic  
lengths, agree with the observed results of potential drops in  
experiments of the first group [3a], [3b].

Moreover, since the ratio between the calculated value of magnetic  
lengths from expementel data of the first group  
$\displaystyle{\frac{(l_B)_{[3a]}}{(l_B)_{[3b]}}} \approx 1.4$ is  
equal to the ratio between their respective filling factors  
$(\displaystyle{\frac{{\nu}_{[3a]}}{\nu_{[3b]}}})^{\frac{1}{2}} =  
(\displaystyle{\frac{4}{2}})^{\frac{1}{2}} \approx 1.4$.
Then it is importent to mention that in this group, where the  
samples  differ almost only with respect to the actual filling  
factors, the ratio between the width of the observed potential drops  
$\displaystyle{\frac{(\Delta A_H)_{(maximum) [3b]}}{(\Delta  
A_H)_{(maximum) [3a]}}} \approx 1.4$ is  equal to the reciprocal  
ratio of the respective filling factors and the respective magnetic  
lengths, i. e.:

\begin{equation}
\displaystyle{\frac{(\Delta A_H)_{(maximum) [3b]}}{(\Delta  
A_H)_{(maximum) [3a]}}} \approx  
(\displaystyle{\frac{{\nu}_{[3a]}}{\nu_{[3b]}}})^{\frac{1}{2}}  
\approx \displaystyle{\frac{(l_B)_{[3a]}}{(l_B)_{[3b]}}}.
\end{equation}

\bigskip
Thus, not only that the observed ratio between potential drops in  
this group agrees with the theoretical ratio between respective  
potential uncertainties, but this agreement proves even the  
exclusive reciprocal dependence of {\it maximal} potential drops  
from the respective filling factor or equivalently from the  
respective magnetic length, if as in the experiments of this group  
other relevant data are almost the same.

Nevertheless, it is possible that under QHE conditions the  
electronic edge current flows, not within the length scale of  
magnetic length, but further within a larger length scale on the  
sample. Then the position uncertainty of electrons which is the  
actual length scale wherein the edge current flow, have its general  
value which is larger than that of the magnetic length of sample  
$\Delta x_H > l_B$. Therefore, the value of uncertainty of Hall  
potential is in this cases less than its maximum value, i. e.  
$(\Delta A_H)_{(general)} < (\Delta A_H)_{(maximum)} =  
\displaystyle{\frac{\hbar}{e l_B}}$. In other words in accord with  
the uncertainty relation potential uncertainties or potential drops  
on QHE samples should be observed on a width which are, in general,  
smaller than the respective $ \displaystyle{\frac{\hbar}{e l_B}}$  
values. This is what is varified in some of the experiments of  
second group in Ref. [3c].

Thus, our theoretical results for the general potential uncertainty  
also agree  with the observed results of the second group in Ref.  
[3c] where they report on potential drops on QHE samples, which is  
less than the respective $\displaystyle{\frac{\hbar}{e l_B}}$  
values.

\medskip\bigskip
Footnotes and references


\begin{thebibliography}{100}

\bibitem{min}
F. Ghaboussi, quant-ph/9702054; cond-math/9703080.

\bibitem{K}
K. v. Klitzing, Physica B 204, 111-116 (1995).

\bibitem{all}


, [3 a] P. F. Fontein, et al., Phys. Rev. B., 43, 12090 (1991).

The relevant QHE data in this report for our calculation are $n =  
5.0 \cdot 10^{15} \, m^{-2}$ and $\nu = 4$. This corresponds, in  
accord to $\nu = 2\pi n l_B ^2$, to a magnetic length value $l_B  
\approx 1.4 \cdot 10^{-2}  \mu m$.

[3 b] W. Dietsche, K. v. Klitzing and K. Ploog, Surf. Sci. 361/362,  
289-292 (1996).

The relevant QHE data in this report for our calculation are
$n = 3.7 \cdot 10^{11} cm^{-2}$ and $\nu = 2$. This corresponds to  
a magnetic length value $l_B \approx 10^{-2} \mu m$.


[3 c] Further reports on potential drops are:

G. Ebert, K. v. Klitzing and G. Weimann, J. Phys. C.: Solid state  
Phys., 18 (1985) L257-L260;

H. Z. Zheng, et al. Phys. Rev. B 32 (1985) 5506;

E. K. Sichel, et al. Phys. Rev. B 32 (1985) 6975;

P. F. Fontein, et al. Surface Scince 263 (1992) 91;

S. Takaoka, et al., Phys. Rev. Lett 72 (1994) 3080.

\bibitem{qml}

, [4a] Landau-Lifschitz, "Quantum Mechanics (Non-Relativistic  
Theory)" Vol.III (Pergamon Press 1987).

[4b] H. Aoki: Rep. Prog. Phys. 50, (1987), 655.

\bibitem{wood}
N. Woodhouse,"Geometric Quantization",  (Clarendon Press, 1980,  
1990) Oxford University.


\bibitem{dim}
Recall that in geometic units, where $(mass \sim  L^{-1}) \ ,  
(length \sim time \sim L)$ and $(e \sim L^0)$, the action $\hbar$ is  
of dimension $L^2$. Hence, $\Delta A \sim  
\displaystyle{\frac{\hbar}{e l_B}}$ is of dimension $L$, which  
agrees with dimension of the observed width of potential drops.
Recall also that, in view of dynamical behaviour of electrons in a  
magnetic field, one should consider the electrodynamical value $(e  
\approx 1.6 \cdot 10^{-19} \; Amper \cdot S)$ in  
$\displaystyle{\frac{\hbar}{e}}$. Then in accord with $\hbar \approx  
1.6 \cdot 10^{-27}$ one has in the (C. G. S.) system  
$\displaystyle{\frac{\hbar}{e}} \approx 10^{-8} \ cm^2 = 1 \ \mu  
m^2$.



\end{thebibliography}
\end{document}